\documentclass{emulateapj}
\slugcomment{To appear in The Astrophysical Journal}

\shortauthors{Halpern et al.}
\shorttitle{Search for Pulsations from 3EG J1835+5918}

\begin{document}

%
%
\def\xray{RX~J1836.2+5925}
\def\source{3EG J1835+5918}
\def\ro{{\it ROSAT\/}}
\def\chandra{{\it Chandra\/}}
\def\glast{{\it GLAST\/}}
\def\hst{{\it Hubble Space Telescope}}

\title{The Next Geminga: Search for Radio and X-ray Pulsations\\
from the Neutron Star Identified with 3EG J1835+5918}

\author{J. P. Halpern, F. Camilo, and E. V. Gotthelf}
\affil{Columbia Astrophysics Laboratory, Columbia University,
New York, NY 10027}

\begin{abstract}
We report unsuccessful searches for pulsations
from the neutron star \xray\ identified
with the EGRET source \source.  
A 24-hr observation with the NRAO Green Bank Telescope at 820 MHz
placed an upper limit on flux density of 17 $\mu$Jy for $P\ga 10$~ms,
and gradually increasing limits for $1\la P\la 10$~ms.  The equivalent
luminosity is lower than that of any known pulsar with the possible
exception of the radio-quiet $\gamma$-ray pulsar Geminga.
A set of observations with the {\it Chandra X-ray Observatory}
HRC totaling 118~ks revealed no pulsar with 1~ms~$\leq P \leq 10$~s.
The upper limit on its pulsed fraction is 35\% assuming a sinusoidal
pulse shape.  The position of \xray\
in \chandra\ observations separated by 3~years is unchanged within errors,
leading to an upper limit on its proper motion of
$<0.14^{\prime\prime}$~yr$^{-1}$, or $v_t<530$ km~s$^{-1}$ at $d = 800$~pc,
a maximum distance estimated from its thermal X-ray spectrum.
With these null results, the properties of \source\ and its
X-ray counterpart \xray\ are consistent with a more distant or older
version of Geminga, or perhaps a recycled pulsar.  Having nearly
exhausted the capabilities of current instrumentation at all wavelengths,
it will likely fall to the {\it Gamma-ray Large Area Space Telescope}
to discover pulsations from \source.
\end{abstract}


\section{Introduction}

The brightest of the ``unidentified'' high-Galactic-latitude EGRET sources
\citep{har99},
\source\ at $(\ell, b)=(89^{\circ},+25^{\circ})$ has long been
associated with the X-ray emitting neutron-star \xray, which
remains its most plausible counterpart even though pulsations
have not been detected at any wavelength.  Unlike blazars,
which are highly variable and have steeper $\gamma$-ray spectra,
\source\ shows no evidence for long-term variability,
and its spectrum can be fitted by a relatively flat power law of
photon index $\Gamma = 1.7$ from 70~MeV to
4~GeV, with a turn-down above 4 GeV \citep{rei01}, similar to
known $\gamma$-ray pulsars.
\xray\ is the only optically undetected X-ray source in the
$\gamma$-ray error box, all the others being classified as unlikely
$\gamma$-ray emitters \citep{mir00}.  The detection of \xray\ as
a weak, ultrasoft source in the \ro\ All-Sky Survey \citep{mir01}
suggested that it is a thermally emitting neutron star that is either
older or more distant than the $\gamma$-ray pulsar Geminga
\citep{hal93,big96}.

Observations using the \chandra\ Advanced CCD Imaging Camera (ACIS)
and the \hst\ ({\it HST\/})
further supported this interpretation \citep{hal02}.
Two components were required to fit the \chandra\ X-ray
spectrum, a blackbody of $T \sim 3.5 \times 10^5$~K
and a power law of photon index $\Gamma = 2.2 \pm 0.6$.  The non-thermal
X-ray component is important evidence of magnetospheric activity that is
found in all $\gamma$-ray pulsars, but not in all cooling neutron stars.
An optical upper limit of $V > 28.5$ from the {\it HST\/}
verified the neutron-star
interpretation of \xray\ from its extreme X-ray-to-optical flux ratio,
$f_X/f_V > 6000$.  We also used the thermal X-ray
spectrum and optical limit to bound the
distance to \xray\ in the range $250 < d < 800$~pc.y
The energetics of \source\ are plausible for a pulsar
at $d <  800$~pc since its $\gamma$-ray luminosity
(assumed isotropic) is $3.8 \times 10^{34}\,(d/800\,{\rm pc})^2$ ergs~s$^{-1}$,
comparable to the spin-down power $\dot E = -I\Omega \dot\Omega$ of Geminga
($3.3 \times 10^{34}$ ergs~s$^{-1}$).  Efficiencies approaching 100\%
are achieved by those $\gamma$-ray pulsars having the smallest
spin-down power.

\citet{hal02} also searched for radio pulsations from \xray\
using the Jodrell Bank Lovell telescope at a frequency
of 1.4~GHz, achieving an upper limit of $S_{1.4} < 0.1$~mJy.
Adopting an upper limit of $d < 800$~pc,
the pulsed pseudo-luminosity limit of \xray\ is
$L_{1.4} \equiv S_{1.4} d^2 < 0.064$~mJy~kpc$^2$.
There are only four pulsars known with $L_{1.4} < 0.1$~mJy~kpc$^2$
\citep{cam02} in addition to Geminga
\citep[][and references therein]{mcl99}.

In this paper, we report an even more sensitive search for radio pulsations
from \xray\ using the NRAO Green Bank Telescope (GBT),
as well as the first X-ray pulsation search with the \chandra\ High
Resolution Camera (HRC-S).  These investigations approach the sensitivity
limits of existing instrumentation to search for the pulsar in \source.

\section{GBT Radio Pulsar Search}

\xray\ is circumpolar at the GBT, where on 2002 December
6 we observed it for 24\,hr using the BCPM spectrometer \citep{bac97}.
Due to the intermediate latitude of the target, with a
relatively small predicted dispersion measure (DM) and low Galactic
background temperature, we observed at a central frequency of 820\,MHz,
with a bandwidth of 48\,MHz in each of two polarizations.  The signals
from corresponding polarization channels were summed in hardware, and
the total-power samples from each of 96 frequency channels were sampled
every $72\,\mu$s and recorded to disk for off-line analysis.

We analyzed the data using standard pulsar search techniques, implemented
in the PRESTO software package \citep{ran02}.  First we identified
and excised radio-frequency interference.  We then dedispersed the data in
the DM range 0--110~cm$^{-3}$~pc, with resolution of 0.3~cm$^{-3}$~pc.
This is twice the maximum Galactic DM predicted for this line of sight by
the \citet{cor02} electron density model; for the distance range
of 250--800~pc, within which the neutron star likely lies (\S 1), the
predicted DM range is 2--9~cm$^{-3}$~pc.  Each time series contained 1.2
billion points, and would have been needlessly challenging to analyze.
We therefore down-sampled the data to a resolution of 0.288~ms,
resulting in much more reasonable time series of 300 million points.
For a DM of 40~cm$^{-3}$~pc, the smearing caused by dispersion across
one of the frequency channels is 0.3~ms, so that for smaller DMs the
effective time resolution of the search was about 0.3~ms.

We performed
the periodicity search using fast-Fourier techniques, and were sensitive
to periods $P \ga 1$~ms.  In order to improve sensitivity for a range of
pulse duty cycles, we incoherently summed up to 16 harmonics of the data;
for $P \ga 10$~ms, we maintained good sensitivity for duty cycles down
to about $0.03P$.  The sensitivity at long periods decreases gradually
due to red noise, but this is not a significant concern for $P \la 5$~s.
Unfortunately, we found no convincing pulsar candidates in this search.

Using the standard modification to the radiometer equation, assuming
a pulse shape with 10\% duty cycle, we were sensitive to pulsars with
$P \ga 10$\,ms having a period-averaged flux density at 820\,MHz $\ga
17\,\mu$Jy, with gradually decreasing sensitivity for smaller periods.
Converting this to the more usual 1.4\,GHz pulsar search frequency
with a typical spectral index of --1.6 \citep{lor95} gives
$S_{1.4} \la 7\ \mu$Jy.  For a distance of 800~pc, this corresponds
$L_{1.4} \la 5\ \mu$Jy~kpc$^2$.
All known radio pulsars have a greater $L_{1.4}$ \citep{cam02}.

Based on these results, it is therefore unlikely that \xray\
is a radio pulsar beaming toward the Earth.  More likely, it is either
beaming away, or is not an active radio pulsar at all.

\section{Chandra HRC Pulsar Search}

\begin{deluxetable}{llrc}
\tablecaption{Chandra HRC Observations of \xray }
\tablehead{
\colhead{ObsID} & \colhead{Start Time} & \colhead{Live Time} & Counts\\
\colhead{}      & \colhead{(UT)}       & \colhead{(s)}       &
}
\startdata
6182            & 2005 Feb 28 19:27    &  45,018             & 284\\
4606            & 2005 Mar 9 15:30     &  28,030             & 189\\
6183            & 2005 Mar 11 08:19    &  44,999             & 256\\
\hline
Total           &                      & 118,047             & 729
\enddata
\label{logtable}
\end{deluxetable}

Three observations of \xray\ totaling 118~ks of live time
were performed between 2005 February 28
and 2005 March 11 using the \chandra\ HRC-S in timing mode, which
has time resolution of $16\ \mu$s, and provides absolute accuracy of
$42\ \mu$s as determined from observations of the Crab
pulsar\footnote{http://cxc.harvard.edu/cal/}.
Table~\ref{logtable} lists the basic parameters of the observations,
including the number of photons extracted from a 
$1.\!^{\prime\prime}2$ radius aperture.
The live time is $99.6\%$ of the elapsed time, limited by event processing.
The total number of photons detected from \xray\
is 729, of which 20 are estimated to be background (using nearby,
source-free regions of the image).  The count rates from
the three observations are within $9\%$ or less of the average
count rate, $0.006$~s$^{-1}$,
consistent with no variation.  The spatial
distribution of photons is consistent with a point source.
Originally conceived as a contiguous, 135~ks
pointing, the observation could not be executed as planned because of
new operational restrictions that came into effect.
The resulting large gaps
in time diminish the sensitivity to pulsations
because the increased number of independent frequencies
and frequency derivatives that have to be searched
yield more false candidates.

The total time span of the observations
from beginning to end is $T=9.6 \times 10^5$~s,
which determines the independent intervals for
frequency and frequency derivative, $\Delta f = 1/2T$
and $\Delta \dot f = 1/T^2$, respectively.  We chose
a frequency step size $\delta f = 2 \times 10^{-7}$~Hz
to oversample the independent frequency interval by a factor
of $\approx 2.5$, and $\delta \dot f = 4 \times 10^{-13}$ Hz~s$^{-1}$
to achieve a similar oversampling factor for frequency derivative.

\begin{figure}
\centerline{
\includegraphics[width=3.3in,angle=0.]{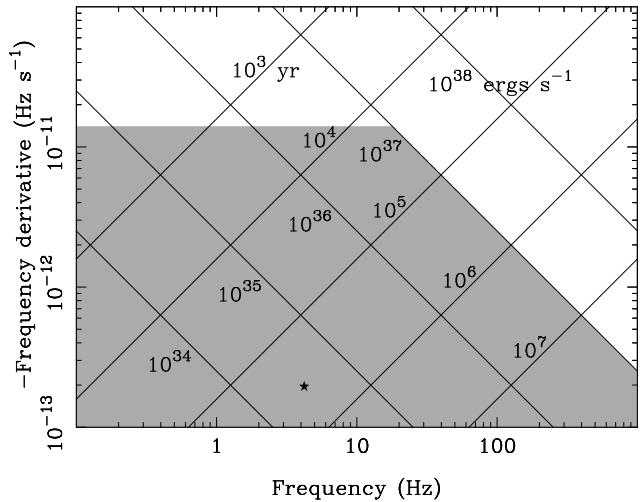}
}
\caption{The parameter space ($f,\dot f$) searched
for X-ray pulsations from \xray\ is shaded.
The step size $\delta \dot f = 4 \times 10^{-13}$
Hz s$^{-1}$, and the search includes $\dot f = 0$.
Lines of constant $\dot E$ and $\tau_c$ are shown.
The star indicates the parameters of Geminga.
}
\label{search}
\end{figure}
  
We transformed the photon arrival times to Barycentric Dynamical Time,
and did coherent searches of the full data set
using the discrete Fourier transform technique, also known
as the Rayleigh or $Z^2_1$ test \citep{buc83}.
We searched the entire range of ($f,\dot f$) parameter space
for a pulsar that has $0.1 \leq f \leq 1000$~Hz,
characteristic age $\tau_c \equiv -f/2\dot f > 20$~kyr,
as well as $\dot E = -4\pi^2If\dot f\leq 10^{37}$ ergs~s$^{-1}$.
In practice, we found it simplest to confine the search to
the region shaded in Figure~\ref{search}, bounded by
$\dot f \leq 1.4 \times 10^{-11}$ Hz~s$^{-1}$
for $f < 17.9$~Hz, and $\dot f \leq 2.5 \times 10^{-10}f^{-1}$ Hz~s$^{-1}$
for $f \geq 17.9$~Hz.
The total number of trials thus defined
is $\approx 1.6 \times 10^{10}$, an oversampling
of the independent trials by a factor of $\approx 7$.

This is a liberal search range that is justified as follows:
\xray\ would have a blackbody temperature
greater than its observed $\sim 3.5 \times 10^5$~K if it
were younger than 20~kyr; an independent
argument for an older age is given in \S 4.
It would have a 
wind nebula or stronger nonthermal X-ray component if
$\dot E > 10^{37}$ ergs s$^{-1}$.  In view of
the distance limit $d < 800$~pc, the ratio
$L_x/\dot E \leq 10^{-6}$ if $\dot E = 10^{37}$ ergs~s$^{-1}$,
a smaller ratio than all other pulsars, which typically
have $10^{-4} < L_x/\dot E < 10^{-2}$.  So it is unlikely
that $\dot E$ is as large as $10^{37}$ ergs~s$^{-1}$.

The range of
parameters searched also includes all of the known
millisecond (recycled) pulsars, and is effective as
long as the pulsar does not have a neutron star binary companion.
The deep {\it HST\/} limit, corresponding to absolute
magnitude $M_V > 19$ at $d=800$~pc, rules out the more
common white dwarf companions.

No significant pulsed signal was detected.  The largest values
of $Z^2_1$ found were $\approx 44$.  The theoretical 
distribution of $Z^2_n$ follows that of $\chi^2$
with $2n$ degrees of freedom.  For $n=1$, the distribution
is an exponential with a mean of 2,
so the single-trial probability that
$Z^2_1 \geq 44$ by chance is $2.8 \times 10^{-10}$.  This
is expected to arise randomly in $\approx 10^{10}$ independent trials.
\citet{lea83} showed that to detect sinusoidal pulsation with
amplitude (pulsed fraction) $f_p$ at a power level $S = Z^2_1$
with 50\% probability,
the number of photons needed is $N=2S/f_p^2$.  In our case, $N = 729$
(709 corrected for background), so
we find $f_p \le 0.35$ corresponding to $Z^2_1 \le 44$.
We verified this analytic expression by examining folded light curves
at the periods associated with the maximum values of $Z^2_1$.

It is not surprising that the pulsed fraction of a $\gamma$-ray pulsar
in X-rays should be less than $35\%$.  This is not a very sensitive
limit consdering that other $\gamma$-ray pulsars
whose soft X-rays are predominantly thermal (Geminga, PSR B1055--52, 
possibly PSR B0656+14) have even smaller pulsed fractions
\citep[30\%, 21\%, and 12\%, respectively,][]{del05}.
In these cases, the pulse profiles are quasi-sinusoidal.
The $Z^2_1$ test is therefore a good one for
\xray, whose X-rays are also dominated by a soft thermal source
within the response of the \chandra\ HRC, although the result is
not very restrictive in this case.
We also note that the Vela pulsar has a more complicated pulse shape,
while its pulsed fraction is only 9\% \citep{man07}.

\section{Chandra Astrometry}

In order to test for proper motion of the neutron star,
we compared the position of \xray\ on the HRC with the ACIS
position obtained 3 years earlier \citep{hal02}.
First, we updated the ACIS position to the USNO-B1.0
system using optically identified sources in the ACIS
image. The result
is R.A. = $18^{\rm h}36^{\rm m}13.\!^{\rm s}685$, 
decl. = $+59^{\circ}25^{\prime}29.\!^{\prime\prime}95$ (J2000.0),
and differs from the previously used USNO-A2.0 system
by $0.\!^{\prime\prime}3$.  Then we registered the HRC image
to the ACIS image using seven X-ray sources in the vicinity of
\xray, which required a zero-point shift of the HRC image of
$+0.\!^{\rm s}052$ in R.A. and $+0.\!^{\prime\prime}89$ in decl..
The resulting HRC position of \xray\ is
R.A. = $18^{\rm h}36^{\rm m}13.\!^{\rm s}674$, 
decl. = $+59^{\circ}25^{\prime}30.\!^{\prime\prime}15$ (J2000.0).

The difference between the positions of \xray\ in 2002 and 2005
is therefore only
$0.\!^{\prime\prime}21$, which is comparable to their combined
statistical errors.  In order to bound the possible proper
motion, we adopt an upper limit that is twice this difference,
or $0.\!^{\prime\prime}42$, corresponding to
$\mu < 0.\!^{\prime\prime}14$~yr$^{-1}$.
This allows the neutron star to have traveled from a birth in
the Galactic plane to its present position in $6.4 \times 10^5$~yr
or longer, which is not an unreasonable age for its X-ray temperature.
At the maximum distance of 800~pc, the limit on $\mu$
corresponds to a tangential velocity limit
$v_t<530$ km~s$^{-1}$, which is typical for radio pulsars.

\section{Discussion and Conclusions}

Even in the absence of detectable pulsations, \xray\ remains the
leading (and only) candidate for identification with \source.
Its soft X-ray spectrum, and absence of optical and radio emission,
support the hypothesis that it is an older and possibly more distant
cousin of the Geminga pulsar.  The upper limit on 
proper motion that we derived is a significant new observational
constraint, allowing an age of at least $6.4 \times 10^5$~yr
for \xray\ if it was born in the Galactic plane.  This compares
favorably with the characteristic age $\tau_c = 3.4 \times 10^5$~yr
of Geminga, and is consistent with a surface temperature of only
$\sim 3.5 \times 10^5$~K, compared with $4.8 \times 10^5$~K for
Geminga \citep{car04,jac05,kar05}.

Assuming that the age and
distance of \xray\ are both larger than those of Geminga,
it is likely that \source\ is a maximally efficient $\gamma$-ray
pulsar close to its death line \citep{che93},
following the trend of $\gamma$-ray efficiency
increasing with decreasing spin-down power \citep{tho97,tho99}.
Alternatively, we cannot rule out that \xray\ is a millisecond
pulsar with similar spin-down power and magnetospheric gap
voltage as Geminga.  If so, its thermal X-ray emission may be due to
surface reheating by a magnetospheric accelerator.

\section{Prospects for GLAST}

It is not now possible to make a significantly more sensitive
search for X-ray pulsations from \xray, as the operational
limitations of \chandra\ no longer allow long, contiguous pointings.
While {\it XMM-Newton} is in principle more sensitive, it has
not been used to observe \xray, which lies near
the Earth-avoidance zone of the satellite orbit, with maximum 
allowed pointing durations of $\approx 20$~ks.

Finally, the upcoming
{\it Gamma-ray Large Area Space Telescope} (\glast)
may be the best hope of detecting
pulsations from \source. {\it GLAST}
will certainly reduce the positional uncertainty of \source,
which will be an independent test of association with \xray,
and additional motivation for a dedicated pulsar search.
For this, a sufficiently long, pointed
observation is best, as argued by \citet{ran07}.

We estimate the expected \glast\ count rate
from \source\ using its photon flux of $5.9 \times 10^{-7}$
cm$^{-2}$~s$^{-1}$ between 100~MeV and 4~GeV \citep{rei01}.
Assuming an average on-axis effective area of 5000~cm$^2$
for the \glast\ Large Area Telescope (LAT) in this energy
range, the predicted count rate is 0.003~s$^{-1}$.
A more detailed calculation, convolving the spectrum of
\source\ from Figure~3 of \citet{rei01} with the LAT on-axis
effective area curve\footnote{http://www-glast.slac.stanford.edu/software/IS/glast\_lat\_performance.htm},
predicts a similar rate of 0.0027~s$^{-1}$ in the 100--4000~MeV range.
This is almost half the count rate from the \chandra\ HRC,
and holds great promise for a pulsar discovery.

It is likely that \glast\ will succeed where \chandra\
failed because, unlike thermal X-rays, $\gamma$-rays 
have sharp peaks with pulsed amplitude approaching 100\%.
In 1 week of elapsed time, including Earth blockage,
\glast\ can collect as many photons from \source\ as 
we have obtained using \chandra.  On-source time 
in a pointed observation
of \source\ can be higher than average because
the source is close to the ecliptic pole.  Also,
Galactic background is negligible at these coordinates
$(\ell, b)=(89^{\circ},+25^{\circ})$.
If \glast\ is successful, the
\chandra\ photons analyzed here may yet be used to find a
coincident period and period derivative with confidence,
thus establishing the identity of \source\ with \xray,
and determining its physical parameters.

\acknowledgements

Support for this work was provided by NASA through {\it Chandra} Award
SAO GO4-5058X issued by the {\it Chandra} X-ray Observatory Center,
which is operated by the Smithsonian Astrophysical Observatory
for and on behalf of NASA under contract NAS8-03060.
The GBT is operated by the National Radio Astronomy Observatory,
a facility of the National Science Foundation operated under cooperative
agreement by Associated Universities, Inc.

\end{document}